\documentclass[11pt,twoside]{article}
\usepackage[latin9]{inputenc}
\usepackage{graphicx}
\usepackage[authoryear]{natbib}

\makeatletter

\usepackage{asp2010}

\resetcounters

\renewcommand\maketitle{}

\makeatother

\begin{document}

\title{Is HDF5 a good format to replace UVFITS?}

\author{Danny~C.~Price$^{1}$, Benjamin R. Barsdell$^{1}$, and Lincoln
J. Greenhill$^{1}$ \affil{$^{1}$Harvard-Smithsonian Center for
Astrophysics, MS 42 60 Garden Street, Cambridge MA 02143} }
\maketitle
\begin{abstract}
The FITS (Flexible Image Transport System) data format was developed
in the late 1970s for storage and exchange of astronomy-related image
data. Since then, it has become a standard file format not only for
images, but also for radio interferometer data (e.g. UVFITS, FITS-IDI).
But is FITS the right format for next-generation telescopes to adopt?
The newer Hierarchical Data Format (HDF5) file format offers considerable
advantages over FITS, but has yet to gain widespread adoption within
radio astronomy. One of the major holdbacks is that HDF5 is not well
supported by data reduction software packages. Here, we present a
comparison of FITS, HDF5, and the MeasurementSet (MS) format for storage
of interferometric data. In addition, we present a tool for converting
between formats. We show that the underlying data model of FITS can
be ported to HDF5, a first step toward achieving wider HDF5 support.
\end{abstract}

\section{Introduction}

The Flexible Image Transport System (FITS) data format is the most
widespread standard for the storage and exchange of datasets within
astronomy. Since its inception \citep{1979ipia.coll..445W,1980SPIE..264..298G},
FITS has enjoyed several decades of widespread usage. The success
of FITS has been attributed in part to the guiding maxim ``once FITS,
always FITS'': that changes to the standard must be incremental
and must not break backward compability. For this reason, it is familiar
to many generations of astronomers, and a large ecosystem of software
 has in turn motivated further adoption of the standard. In particular,
the \texttt{CFITSIO} library (ascl:1010.001)  for the reading and
writing of FITS files has become the \emph{de facto} standard.

FITS has necessarily evolved over the years, with the addition of
features such as random groups \citep{1981A&AS...44..371G}, tables
\citep{Harten1998,Cotton1995}, and compression \citep{2002SPIE.4847..444P};
FITS is now officially at version 3.0 \citep{2010A&A...524A..42P}.
However, these changes have been relatively minor iterations upon
the core FITS format.

The limitiations of FITS have been previously acknowledged and documented,
for example in \citet{2014ASPC..485..351T} and \citeauthor{Thomas:2014vq}
(in press). In \citet{Kitaeff:2014ua}, the authors consider JPEG2000
as an alternative format for astronomical images. \citet{2001ASPC..238..487T}
discuss advantages of converting FITS files to XML; \citet{Jennings1995}
considered HDF4 as a format. Here, we consider the immediate, practicable
advantages of HDF5 (Hierarchical Data Format) as a data storage format
for visibility data. We show that data in FITS format can be converted
in a straightforward fashion to HDF5 format, and that conventions
for the storage of visibility data can be ported to HDF5.

\subsection{Definitions}

In order to discuss data storage methods and file formats without
ambiguity, we first need to clarify our vocabulary:
\begin{itemize}
\item \emph{Data model:} a high-level, conceptual model of data, types of
data, and how data are organized, e.g. ``group'' and ``dataset''.
\item \emph{Data schema: }a lower-level, domain-specific ontology (i.e.
framework that gives meaning) of how data and metadata are arranged
inside a data model. 
\item \emph{Storage model:} how objects from the data model are mapped to
bytes within an address space on storage media. 
\item \emph{Convention: }a documented data schema that has widespread acceptance
within a community of users.
\item \emph{Standard: }the acknowledged, formal specification of a file
format. A standard may or may not define acceptable data models and
schema.
\end{itemize}
From this view, the data model can be seen as \emph{syntax}, while
the data schema may be seen as the ontology that gives \emph{semantics}.
 Neither the FITS nor HDF5 standards define data schema; however
there are registered FITS conventions%
\footnote{\url{http://fits.gsfc.nasa.gov/fits_registry.html}%
} for certain classes of data.

\section{Conventions for visibility data storage}

\subsection{FITS-IDI and UVFITS}

There are two registered FITS conventions for the storage of visibility
data from synthesis imaging radio telescopes: FITS-IDI \citep{Griesen:2008tn},
and UVFITS \citep{Greisen:2012tw}. Both these formats store not only
the visiblility data, but also metadata such as antenna positions,
information on observation setup, and calibration tables.

The two visibility data conventions share many keywords and unit definitions,
but their schema and underlying storage models differ. In UVFITS,
the visibility data are stored in a random group HDU (header data
unit), whereas in FITS-IDI data are stored in a binary table HDU.
In both formats, each row of the table contains columns for the timestamp
and a baseline identifier, along with the multidimensional visibility
array for that timestamp and baseline.

\subsection{CASA MeasurementSets}

An alternative file format for visibility data is the MeasurementSet
(MS; \citealp{Kemball:2001um}). This format is used by the CASA reduction
package (ascl:1107.013) and related software. The storage model for
the MS format is a directory consisting of several data files nested
inside child directories. Unlike FITS and HDF5, MS has no in-built
data compression capability. Visibility data are stored in the MAIN
table; this also provides keys to the various subtables that provide
instrument and observation metadata. The MS standard defines data
schema for images, visibility data and single-dish data. The MS data
model differ significantly to FITS, meaning that conversion requires
a non-trivial mapping. For example, MS files have no equivalent to
the FITS HDU, and while FITS-IDI assigns an integer ID for every baseline,
MS uses a pair of antenna IDs.

\subsection{HDF5 for visibility storage}

There are several advantages of the HDF5 format over both FITS and
MS, the most compelling of these are improvements to the storage model
that are beneficial for large datasets. For example, HDF5 provides
parallel and network I/O, data chunking methods, external (i.e. distributed)
object storage, and a filter pipeline for data compression. Of specific
interest for visibility data is  \texttt{bitshuffle}%
\footnote{\url{https://github.com/kiyo-masui/bitshuffle}%
}, an HDF5 filter designed for fast compression of visibility data.
Using \texttt{bitshuffle} on a 1.2 GB test dataset of data from the
LEDA correlator \citep{Kocz:2014jr}, we achieved lossless compression
ratio of 1.65x, with total file compression and write time of 7.5
s; in comparison the data compressed by 1.40x in 53.0 s using standard
\texttt{gzip}.

HDF5 is already in limited use within astronomy, but no general convention
for visibility data storage currently exists. The LOFAR radio telescope
has developed conventions specifically for storage of LOFAR data \citep{2011ASPC..442..663W,2012ASPC..461..283A},
and the (unfunded) AstroHDF effort sought to explore HDF5 for astronomical
datasets further \citep{2012ASPC..461..871M}.

\section{Conversion between FITS and HDF5}

\begin{figure}
\begin{centering}
\includegraphics[width=1\columnwidth]{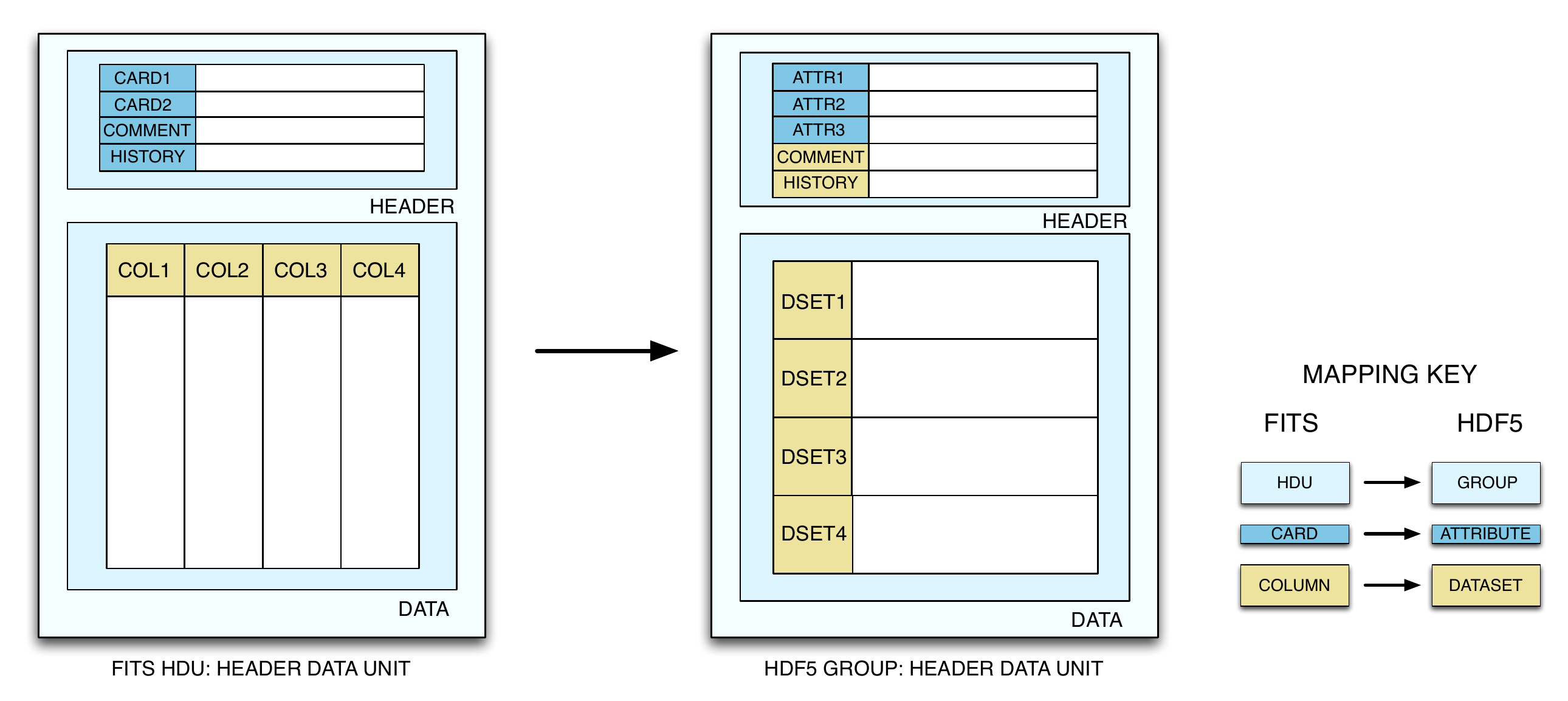}
\par\end{centering}

\protect\caption{Diagram showing the mapping of the FITS data structures into the HDF5
data model. }

\end{figure}

While there are many possible mappings, one of the simplest is to
map the FITS HDU structure to a HDF5 \emph{group}, FITS cards to HDF5
\emph{attributes}, and the FITS data payloads to HDF5 \emph{datasets}
(Fig 1). A similar approach is being undetaken to port the N-dimensional
data format (NDF) to HDF5 (\citeauthor{Jenness:2014wy}, in press).
The hierarchical nature of HDF5 allows the header unit and\emph{ }data
unit within the FITS HDU to be mapped to HDF5 groups within a parent
group; similarly, in the case of FITS tables, each column can be mapped
to a separate dataset\emph{ }within a parent group. All FITS data
types (e.g. float32) have HDF5 equivalents.

We have implemented a utility called \texttt{fits2hdf}\emph{ }that
uses this mapping to convert FITS files into HDF5 format%
\footnote{\url{https://github.com/telegraphic/fits2hdf}%
}. This utility is written in Python, and uses the PyFITS (ascl:1207.009)
and h5py libraries for file I/O. These ``HDFITS'' files have an
additional attribute in the root to identify them as having a HDFITS
data model. While \texttt{fits2hdf} was designed to port UVFITS/FITS-IDI
data into HDF5, any valid FITS file may be converted by this utility.
As the HDFITS data model is a restricted subset of the complete HDF5
data model, any HDFITS file may be converted back into a FITS file
without complication. 

A similar approach can be taken with MS files, converting them into
``HDF-MS'' data models within HDF5. This functionality is provided
by a separate utility, \texttt{ms2hdf}, provided as part of the \texttt{fits2hdf}
package.

\section{Conclusions }

By porting the FITS and MS data models to HDF5, we may begin to leverage
the advantages of HDF5. Doing so maintains familiarity to users and
allows conversion back into FITS/MS. This also limits the amount of
modification required for existing programs to read, write and understand
HDF5-stored data. 

\acknowledgements This work was supported by NSF grants OIA­-1125087,
AST-1106059, and OCI­-1060067. 

\bibliographystyle{asp2010}
\bibliography{hdfidi}

\end{document}